\documentclass[sigconf]{acmart}

\usepackage{multirow}
\usepackage{subfigure}
\usepackage[utf8]{inputenc}
\usepackage{graphicx}
\usepackage{amsmath}
\usepackage{algorithm}
\usepackage{algorithmic}
\usepackage{CJKutf8}
\usepackage{balance}
\usepackage[skip=2.5pt]{caption}
\usepackage{array}
\usepackage{booktabs}
\usepackage{threeparttable}
\usepackage{soul}

\AtBeginDocument{%
  \providecommand\BibTeX{{%
    \normalfont B\kern-0.5em{\scshape i\kern-0.25em b}\kern-0.8em\TeX}}
}

\acmConference[Conference acronym 'XX]{Make sure to enter the correct
  conference title from your rights confirmation emai}{June 03--05,
  2018}{Woodstock, NY}
\copyrightyear{2023} 
\acmYear{2023} 
\setcopyright{acmlicensed}\acmConference[WWW '23 Companion]{Companion Proceedings of the ACM Web Conference 2023}{April 30-May 4, 2023}{Austin, TX, USA}
\acmBooktitle{Companion Proceedings of the ACM Web Conference 2023 (WWW '23 Companion), April 30-May 4, 2023, Austin, TX, USA}
\acmPrice{15.00}
\acmDOI{10.1145/3543873.3584639}
\acmISBN{978-1-4503-9419-2/23/04}

\begin{document}

\title{
    A Multi-Granularity Matching Attention Network for Query Intent Classification in E-commerce Retrieval
}

\author{Chunyuan Yuan\textsuperscript{\rm *}, Yiming Qiu, Mingming Li, Haiqing Hu,  Songlin Wang, Sulong Xu}

\email{
    chunyuany93@outlook.com, {qiuyiming3,limingming65,huhaiqing1,wangsonglin7,xusulong}@jd.com
}
\affiliation{
  \institution{JD.com, Beijing \country{China} } 
}



\begin{abstract}
Query\let\thefootnote\relax\footnotetext{* Corresponding author.} intent classification, which aims at assisting customers to find desired products, has become an essential component of the e-commerce search. Existing query intent classification models either design more exquisite models to enhance the representation learning of queries or explore label-graph and multi-task to facilitate models to learn external information. However, these models cannot capture multi-granularity matching features from queries and categories, which makes them hard to mitigate the gap in the expression between informal queries and categories. 

This paper proposes a \textbf{M}ulti-granularity \textbf{M}atching \textbf{A}ttention  \textbf{N}etwork (MMAN), which contains three modules: a self-matching module, a char-level matching module, and a semantic-level matching module to comprehensively extract features from the query and a query-category interaction matrix. In this way, the model can eliminate the difference in expression between queries and categories for query intent classification. We conduct extensive offline and online A/B experiments, and the results show that the MMAN significantly outperforms the strong baselines, which shows the superiority and effectiveness of MMAN. MMAN has been deployed in production and brings great commercial value for our company.
\end{abstract}

\ccsdesc[500]{Information systems~Query intent}
\ccsdesc[500]{Computing methodologies~Natural language processing}

\keywords{Query intent classification; multi-granularity matching attention network; multi-label text classification; e-commerce retrieval}



\maketitle

\section{Introduction}
Online shopping has become a way of life for people in the last few decades. More and more e-commerce platforms (e.g., eBay, Amazon, Taobao, and JD) provide consumers with hundreds of millions of products. 
Due to the diversity of user needs and commodity types, e-commerce search systems need to have the ability to identify the intention of the search query. 

Query intent classification has attracted increasing attention from both academic and industrial areas. Multi-label classification models such as XML-CNN~\cite{liu2017deep}, LSAN~\cite{xiao2019label}, and HiAGM~\cite{zhou2020hierarchy} have been proposed to learn the contextual information of documents to enhance the representation learning of queries. However, different from the general search engine, the queries of e-commerce applications are usually very short and lack enough contextual signals, and are even insensitive to the word order. These problems make these models cannot capture the most important information of the query. Moreover, a query may have different meanings. For example, the query ``book hotel'' expects the service about getting accommodation. However, the character ``book'' may mislead the model to predict the paper book. These polysemous queries increase the difficulty of precisely classifying the user's intent. 

Some recent query intent classification models, such as CNN~\cite{hashemi2016query,yu2018diverse,zhang2019improving}, LSTM~\cite{yang2016hierarchical,sreelakshmi2018deep,wu2019learning}, and attention-based  models~\cite{chen2019bert,cai2021slim,zhang2021modeling,qiu2022pre}, explore using the correlation between query intent and textual similarity or label-graph to facilitate models to learn external information. Unfortunately, these models are heavily dependent on the training data. It makes these models hard to generalize for the long tail queries that lack enough supervision signal from users.

To concurrently address the above challenges, 
we proposed a \textbf{M}ulti-granularity \textbf{M}atching  \textbf{A}ttention  \textbf{N}etwork (MMAN), which contains three modules: self-matching module, char-level matching module, and semantic-level matching module to comprehensively extract features from the query itself and query-category interaction matrix for mitigating the gap in expression between queries and categories for long-tail query intent classification.

The contributions of this paper can be summarized as follows:
\begin{itemize}
\item We propose a novel strategy that explicitly extends category information to reduce the expression gap between queries and categories.

\item We design an effective model MMAN that contains three major components: self-matching module, char-level matching module, and semantic-level matching module, which focus on query representation learning, long-tail query enhancement, and semantic disambiguation respectively.

\item We conduct extensive offline experiments on two large-scale real-world datasets and online A/B test experiments. Experimental results show that our model achieves significant improvement over state-of-the-art models.
\end{itemize}

\section{Model}
\label{sec:Model}
\begin{figure}[H]
    \vspace*{-0.6\baselineskip}
	\centering
	\includegraphics[scale=0.37]{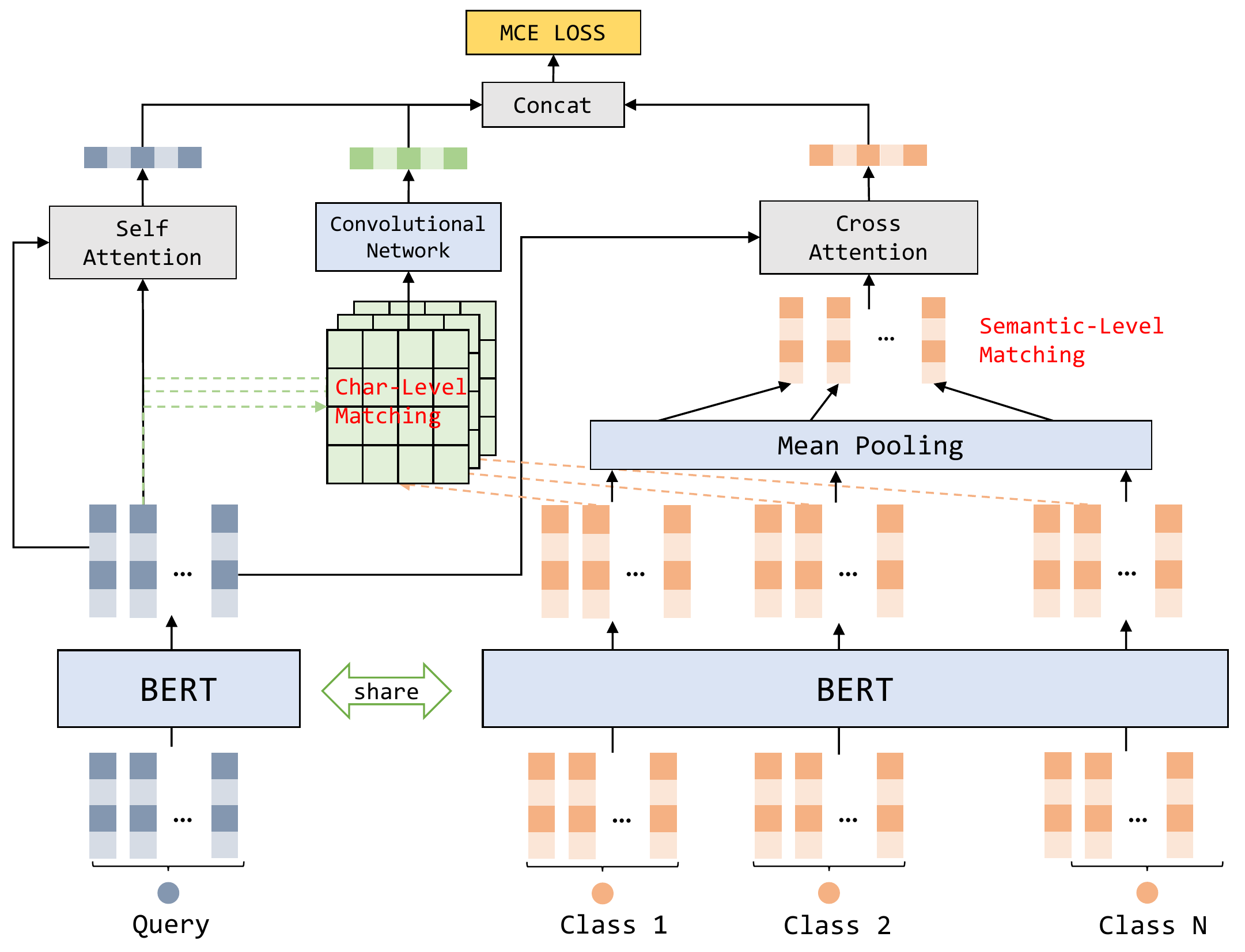}
	\caption{Multi-granularity Matching Attention Network. }
	\label{model_structure}
	\vspace*{-1\baselineskip}
\end{figure}

Figure~\ref{model_structure} illustrates the components of the proposed model, which is mainly composed of four major modules: (1) query and category representation learning module; (2) self-matching module; (3) char-level matching module; and (4) semantic-level matching module. 

\subsection{Query and Category Representation}
The query and category representation is the basis of aligning both into the same semantic space. Pre-trained language models~\cite{kenton2019bert,zhang2019ernie,liu2019roberta} have been widely applied in industrial applications, so we use BERT as the encoder for both the query and categories. 

To better learn the semantics of the product categories, the category character sequence is composed of two parts: (1) category name $n = [n_1, n_2, \ldots, n_{L_n}]$ where ${L_n}$ denotes the number of categories; (2) the selected core product words $m = [m_1, m_2, \ldots, m_{L_m}]$ where ${L_m}$ denotes the number of product words. 



After obtaining the above high-quality product words, we concatenate them with category names and then feed them into BERT to encode category representation. To map queries and categories into the same semantic space, query and category share the same BERT model, which can be formulated as follows:
\begin{equation}
\begin{split}
& \mathbf{Q}_i = \mathrm{BERT_{Token}}([x_1, x_2, \ldots, x_{L_q}])   \,, \\
& \mathbf{C}_j = \mathrm{BERT_{Token}}([n_1, n_2, \ldots, n_{L_n}, m_1, m_2, \ldots, m_{L_m}]) \,, \\
\end{split}
\end{equation}
where $ \mathrm{BERT_{Token}}$ is the final layer of BERT without ``CLS''; $\mathbf{Q}_i \in \mathbb{R}^{L_q \times d}$ and $\mathbf{C}_j \in \mathbb{R}^{L_c \times d} $  denote the token embedding matrix of query and category, respectively.

\subsection{Self-matching module}
A typical text classification model is usually built on pure query text. Thus we leverage this advantage in self-attention style.

Since query words contribute to the query representation differently, we apply the attention mechanism to summarize the query embedding matrix to extract intent-related words that are important to represent the query: 
\begin{equation}
\begin{split}
& \mathbf{u}_i = \mathbf{v}_i\tanh\left(\mathbf{W}_q \mathbf{Q}^T_i  \right) \,,  \\
& \mathbf{q}_i = \sum_{t=1}^{L_q} \mathbf{Q}_{i,t} \mathbf{softmax}(\mathbf{u}_{i,t})  \,,  \\
\end{split}
\end{equation}
where $\mathbf{v}_i \in \mathbb{R}^{1 \times d}$, $\mathbf{W}_q \in \mathbb{R}^{d \times d} $ and $\mathbf{\alpha} = \mathbf{softmax}(\mathbf{u}_{i,t})$ is a score function which determines the importance of words for composing sentence representation about the current query. 


\subsection{Char-level matching module}
For long-tail queries, it lacks enough training samples for the model to precisely predict the user's intent. In this situation, auxiliary knowledge such as category names and core product words are a necessary complement to facilitate the model's decision. 


In order to extract the fine-grained interactive features between the query and all categories, we perform dot-product between query representation $\mathbf{Q}_i$ and category representations $\mathbf{C}_j$,  and then stack each query-category interaction matrix on channel dimension, which can be formulated  as follows:
\begin{equation}
\begin{split}
& \mathbf{M}_j = \mathbf{Q}_i \mathbf{W}_{qc} \mathbf{C}^T_j  \,, \\
& \mathbf{M} = [\mathbf{M}_1, \mathbf{M}_2, \ldots, \mathbf{M}_C]  \,,  \\
\end{split}
\end{equation}
where $W_{qc} \in \mathbb{R}^{d \times d}$ is a trainable weight, $\mathbf{M}_j \in \mathbb{R}^{L_q \times L_c}$ is the interaction matrix between query and category, and $\mathbf{M} \in \mathbb{R}^{|C| \times L_q \times L_c}$ is the feature map on the character level interaction between query and each category.

Then, to extract the fine-grained matching features from the feature map $\mathbf{M}$, we employ a 2D convolution module with a window size $r_w \times r_h$ as follows:
\begin{equation}
\begin{split}
\mathbf{s}^{(k)}_{i,j} = \mathop{ReLU} \left( \sum_{a=0}^{r_w} \sum_{b=0}^{r_h} \mathbf{W}_{a,b} \mathbf{M}^{(k)}_{i+a, j+b} + \mathbf{b} \right)  \,,
\end{split}
\end{equation}
where $k$ denotes the $k$-th channel of feature map $\mathbf{M}$, $\mathbf{W}_{a,b}$ is a convolutional kernel, $\mathbf{b}$ is a bias vector.  We conduct convolution operation on each channel of the feature map $\mathbf{M}$.   
     
Next, a 2D max pooling layer is applied to distill most import features from the feature map $\mathbf{s}$ and can be formulated as follow:
\begin{equation}
\begin{split}
\mathbf{\widetilde{s}}^{(k)}_{i,j} = \mathop{max}_{0 \le c \le p_w} \mathop{max}_{0 \le d \le p_h} \mathbf{s}^{(k)}_{i+c, j+d}   \,,
\end{split}
\end{equation}
where $p_w$ and $p_h$ are the width and height of the 2D max pooling.

Finally, the output of the final feature maps is flattened and mapped into a low dimensional space by a linear transformation layer, denoted as   $\mathbf{Z}_1 \in \mathbb{R}^{|C| \times d}$, which contains fine-grained matching features between query and each category.

\subsection{Semantic-level matching module}
Literal matching features may not enough to fully capture the user's real intent since the word in the query may be polysemous. For example, both ``apple watch'' and ``apple juice'' has high relevance score with ``apple'' on character level, however, semantically different from each other. Therefore, it is also necessary to capture the relevance between query and category on the semantic level.  

To that end, we first obtain the category representation on the semantic level. In this phase, we employ a mean pooling on the time step of the character sequence of category representation $\mathbf{C}_i$, and stack each category representation together:
\begin{equation}
\begin{split}
& \mathbf{c}_i = \mathbf{mean}(\mathbf{C}_i)  \,,  \\
& \mathbf{C} = [\mathbf{c}_1, \mathbf{c}_2, \ldots, \mathbf{c}_{|C|}]  \,,  \\
\end{split}
\end{equation}
where $\mathbf{C} \in \mathbb{R}^{|C| \times d}$ is the representation of all categories.

Then, a cross-attention layer is applied to integrate the query with all categories' representations:
\begin{equation}
\mathbf{Z}_2 = \mathbf{Q}^{T}_{i} \mathbf{softmax}(\mathbf{C} \mathbf{W}_{qs} \mathbf{Q}^T_i)   \,,
\end{equation}
where $\mathbf{W}_{qs} \in \mathbb{R}^{d \times d}$ is a trainable weight, and $\mathbf{Z}_2 \in \mathbb{R}^{|C| \times d}$ is the matching features between query and labels at the semantic level.

\subsection{Training and Inference}
After the above process, we have obtained query self-representation $\mathbf{q}_i$, fine-grained query-category matching features $\mathbf{Z}_1$ and coarse-grained matching features $\mathbf{Z}_2$. All of these representations contribute to predicting the user's intent, and thus we employ matrix multiplication to fuse them together.  Specifically, we introduce the nonlinear transformation layer which is defined as:
\begin{equation}
    \widehat{y} = \mathbf{W}^T_x ReLU \left(\mathbf{q}_i \mathbf{W}_{qf} + [\mathbf{Z}_1, \mathbf{Z}_2] \mathbf{W}_z \right) \,,
\end{equation}
where $\mathbf{W}_{qf} \in \mathbb{R}^{d \times |C|}$, $\mathbf{W}_z \in \mathbb{R}^{2d}$, and $\mathbf{W}_x \in \mathbb{R}^{|C| \times |C|}$ are linear transformation matrices.

In this paper, we use $y \in \mathcal{R}^{|C|}$ to represent the ground-truth label of a query, where $y_i = {0, 1}$ denotes whether the query belongs to the category $i$. The whole framework is trained with the multi-label cross-entropy loss which can be formulated as follows:
\begin{align}
\mathcal{L} &= - \sum ^{C}_{c=1}  y^{c} \log \left( \sigma \left( {\widehat {y}}^c \right) \right) + \left( 1-y^c \right) \log \left( 1 - \sigma \left( \widehat {y}^c \right) \right) \,,
\end{align}
where $\sigma$ is the sigmoid function.

\section{Experiment}
\label{sec:Experiment}

\subsection{Dataset}
\label{sec:Dataset}
In order to verify the effectiveness and generality of the MMAN, we conduct experiments on two large-scale real-world datasets collected from users' click logs of the JD application. The statistics of the datasets are listed in Table~\ref{tab:datset}.

\begin{itemize}
    \item \textbf{Category Data}: To evaluate the performance of MMAN, we randomly sample queries and corresponding clicked products from search logs over a period of one month. The product's category is treated as the query's intent. To filter unreliable categories, we normalized the click frequency of the category and compute the cumulative distribution function (CDF) of the category's probability. When $CDF > 0.9$, the rest with low probabilities are removed. 
    
    \item \textbf{Scene Data}: We divide the collected categories into eight different domains, such as "travel", "hotel booking", "medical consultation", "car service", etc. The categories of the query are mapped into domains, which form the Scene data. Different from the training data, the test dataset is annotated by experts in each domain.  It is worth noting that the experts not only annotate the domain of the query but also annotate all categories to which the query belongs.
\end{itemize}

\begin{table}[t]
    \centering
	\vspace*{-0.5\baselineskip}
    \caption{Dataset statistics.}
	\setlength{\tabcolsep}{2mm}{
        \begin{tabular}{c|cc |cc}
            \toprule
            \multirow{2}{*}{\textbf{Statistic}} &
            \multicolumn{2}{c|}{\textbf{Scene Data}} & \multicolumn{2}{c}{\textbf{Category Data}} \\
            &\textbf{Train} &\textbf{Test} 
            &\textbf{Train} &\textbf{Test}   \\ 
            \hline
            \hline
            \ Queries  & 4,459,214 & 9,877 & 4,593,037 & 9,877 \\
            \hline
            \ Total Labels  & 8 & 8 & 90 & 90 \\
            \hline
            \ Avg. chars & 7.63 & 5.00 & 7.69 & 5.00   \\
            \hline
            \ Avg. \# of labels  & 1.04  & 1.67 & 1.19 & 1.77  \\
            \hline
            \ Min. \# of labels & 1 & 1 & 1 & 1  \\
            \hline
            \ Max. \# of labels & 7 & 3 & 26 & 21  \\
            \bottomrule
        \end{tabular}
    }
    \label{tab:datset}
	\vspace*{-1\baselineskip}
\end{table}

\subsection{Baseline Models}
\label{sec:Baseline Models}
We compare MMAN with several strong baselines, including widely-used multi-label classification methods. The detailed introductions are listed as follows:

(1) Multi-label text classification baselines: \textbf{RCNN}~\cite{lai2015recurrent}: It captures contextual information with the recurrent and convolutional structure for text classification. \textbf{XML-CNN}~\cite{liu2017deep}: It is a CNN-based model, which combines the strengths of CNN models and goes beyond the multi-label co-occurrence patterns. \textbf{LEAM}~\cite{wang2018joint}: It is a label-embedding attentive model, which embeds the words and labels in the same space, and measures the compatibility of word-label pairs. \textbf{LSAN}~\cite{xiao2019label}: It is a label-specific attention network to use document and label text to learn the label-specific document representation with the aid of self- and label-attention mechanisms.

(2) Query intent classification baselines: \textbf{PHC}~\cite{zhang2019improving}: It investigates the correlation between query intent classification and textual similarity and proposes a multi-task framework to optimize both tasks. \textbf{DPHA}~\cite{zhao2019dynamic}: It contains a label-graph-based neural network and a soft training mechanism with correlation-based label representation. \textbf{BERT}~\cite{kenton2019bert}: We use the pre-trained BERT-base~\footnote{https://tfhub.dev/tensorflow/bert\_zh\_L-12\_H-768\_A-12/4} delivered by google, and finetune it on the training set to predict the user's intent. \textbf{SSA-AC}~\cite{zhang2021modeling}: It is an across-context attention model to extract external information from variant queries for intent classification.

\subsection{Experiment Settings}
\label{sec:Experiment Settings}
Following the settings of previous work ~\cite{zhao2019dynamic,zhang2021modeling}, we report the micro and macro precision, recall, and F1-score of the models as the metrics of query intent classification. 

We implement the models based on Tensorflow 2.4. Word embeddings were initialized by pre-trained word2vec~\cite{mikolov2013distributed}, and whose dimensionality is 768. We use two convolution layers to extract char-level feature maps. The convolution layer has 8 [3,3] filters with [1,1] stride, and its max-pooling size is [2,2] with [2,2] stride. We use Adam algorithm~\cite{kingma2014adam} with learning rate as $5e^{-5}$. The max length of the query is set to 16. The threshold of labels is set to 0.5.

\begin{table*}[!htbp]
  \caption{
        The experimental results that compared with multi-label classification and query intent classification models.  
  }
  
  \label{tab:experiment}
  \setlength{\tabcolsep}{2mm}{
      \begin{tabular}{c|ccc|ccc|ccc|ccc}
                \toprule
                \multirow{3}{*}{\textbf{Models}}  & 
                
                \multicolumn{6}{c|}{\textbf{Scene Data}} & \multicolumn{6}{c}{\textbf{Category Data}} \\
                
                & \multicolumn{3}{c|}{\textbf{Micro}}  
                & \multicolumn{3}{c|}{\textbf{Macro}} 
                & \multicolumn{3}{c|}{\textbf{Micro}}  
                & \multicolumn{3}{c}{\textbf{Macro}}  \\
                
                &\textbf{Prec.} &\textbf{Recall} &\textbf{F1}
                &\textbf{Prec.} &\textbf{Recall} &\textbf{F1}
                &\textbf{Prec.} &\textbf{Recall} &\textbf{F1}
                &\textbf{Prec.} &\textbf{Recall} &\textbf{F1} \\
                \midrule
                \midrule
                RCNN~\cite{lai2015recurrent}      & 94.14  & 77.67 & 85.11  & 83.09  & 86.01  & 83.69  & 69.76  & 54.03 & 60.89  & 70.51  & 62.42 & 62.15 \\
                XML-CNN~\cite{liu2017deep}   & 94.73  & 76.00 & 84.34    & 80.87  & 86.47 & 81.91  & 66.73  & 56.36 & 61.11  & 68.08  & 64.15 & 62.12 \\
                LEAM~\cite{wang2018joint}      & 94.19  & 68.46 & 79.29  & 88.84  & 78.60 & 82.84 & 72.67  & 49.91  & 59.18  & 69.96  & 47.56 & 52.15  \\
                LSAN~\cite{xiao2019label}      & 94.73  & 74.14 & 83.18  & 80.31  & 86.05 & 81.48  & 68.33  & 51.36 & 58.64  & 71.64  & 61.00 & 61.93 \\
                
                \midrule
                
                PHC~\cite{zhang2019improving}       & 94.63  & 77.93 & 85.47  & 83.17  & 86.62 & 83.74  & 60.12  & \textbf{59.41} & 59.76  & 64.08  & 64.90  & 60.67  \\

                DPHA~\cite{zhao2019dynamic}      & 95.23  & 77.43 & 85.41  & 82.01  & 84.35  & 82.06  & 71.55  & 54.06 & 61.58  & 75.39  & 54.99  & 61.83  \\

                SSA-AC~\cite{zhang2021modeling}   & 94.82  & 78.15 & 85.68  & 84.15 & 84.26 & 83.92  & 72.36  & 53.20  & 61.32  & 74.38  & 62.19 & 63.38 \\
                \midrule
                \textbf{MMAN}  & \textbf{95.52}  & \textbf{82.26} & \textbf{88.39}  & \textbf{87.26}  & 86.15 & \textbf{85.93}    & \textbf{75.64}  & 55.07  & \textbf{63.74}  & \textbf{75.77}  & 64.56 & \textbf{66.47}   \\
                \ w/o self-matching  & \textbf{96.03}  & 81.24 & 88.02  & \textbf{88.14}  & 85.72 & 84.86    & 75.25  & 54.35 & 63.11  & 73.26  & 64.08 & 65.68  \\
                \ w/o char matching  & 95.16  & 80.28 & 87.09  & 82.12  & \textbf{89.38} & 83.74    & 68.72  & \textbf{57.13} & 62.39  & 72.16  & 62.58 & 65.12  \\
                \ w/o semantic matching  & 95.86  & 81.14 & 87.89  & 84.36  & 87.62 & 84.15   & 72.18  & 56.16 & 63.17  & 73.61  & 63.27 & 65.05  \\
                
                BERT~\cite{kenton2019bert}   & 95.39  & 79.22 & 86.56  & 81.20 & 88.48 & 83.00 & 65.88  & 56.23  & 60.67  & 68.47  & \textbf{67.28} & 64.53   \\
                \bottomrule
        \end{tabular}
    }
    \vspace*{-0.5\baselineskip}
\end{table*}

\subsection{Experimental Results and Analysis}
The experimental results are shown in Table~\ref{tab:experiment}. The results indicate that MMAN outperforms all baselines on two large-scale real-world datasets. Specifically, we have the following observations:

(1) MMAN significantly outperforms the multi-label classification baselines. A similar phenomenon can be observed in the comparison between query intent classification and multi-label classification models. Actually, most of these models are more suitable for contextual modeling of long texts. For queries typed by users on e-commerce, most of them are very short and lack contextual information, and are even insensitive to word order. These problems make them hard to capture the most important information for users' real intent. 

(2) Compared with recent intent classification methods, MMAN achieves better performance on both datasets. Although these models also utilize the interaction information between query and labels, all of them only consider the matching features at the semantic level, which makes the models unable to mitigate the gap between informal query expression and formal category information and hard to capture the fine-grained clues for long tail classification.

(3) After removing these three modules of MMAN, we can see that the micro and macro F1 decay about 3\% compared with the complete model. This result further shows that all of these components in MMAN provide complementary information to each other, and are requisite for intent classification.

In conclusion, MMAN achieves significant improvement over these strong baselines on the metrics of micro-F1 and macro-F1, which further confirms the effectiveness of learning multi-granularity matching features from query and category interaction, which facilitates the model to mitigate the gap of expressions between query and category and learns the most significant features for capturing the user's real intent.

\subsection{Online Evaluation}
Before being launched in production, we routinely deploy the MMAN on the JD search engine and make it randomly serve 10\% traffic as the test group. During the A/B testing period, we monitor the performance of MMAN and compare it with the previously deployed model. This period conventionally lasts for at least a week. For online evaluation, we use some business metrics: Page Views (PV), Product Clicks (Click), Gross Merchandise Volume (GMV), UV value, and Conversion Rate of Users (UCVR).

\begin{table}[t]
  \centering
  \caption{Online improvements of the MMAN. Improvements are statistically significant with $p < 0.05$  on paired t-test.}
  \label{online_performance_uv_ucvr}
  \setlength{\tabcolsep}{2.5mm}{
  \begin{tabular}{c|cccc}
            \toprule
            &\textbf{GMV} &\textbf{UV value} &\textbf{UCVR}
            \\
            \hline
            \ Online model (BERT) & - & - & -    \\
            \ MMAN  & +0.351\% & +0.401\% & +0.113\%   \\
            \bottomrule
    \end{tabular}}
            \vspace*{-0.5\baselineskip}
\end{table}

\begin{table}[t]
  \centering
  \caption{Online performance of the MMAN compared with the online BERT model. Improvements of MMAN are statistically significant with $p < 0.01$  on paired t-test. }
  \label{online_performance_pv_click}
  \setlength{\tabcolsep}{3mm}{
  \begin{tabular}{c|cccc}
            \toprule
            Scene &\textbf{PV} &\textbf{Click}
            \\
            \hline
            hotel booking & +74.85\% & +36.49\%  \\
            travel and vacation   & +9.94\% & +1.75\%  \\
            checkup service  & +5.95\% & +4.89\%  \\
            aesthetic medicine  & +44.42\% & +26.30\%  \\
            medical consultation  & +22.76\% & +5.66\%  \\
            car service   & +20.17\% & +9.38\%  \\
            furniture customization  & +10.07\% & +6.60\%  \\
            Overall   &+19.62\%   & +7.78\%  \\
            \bottomrule
    \end{tabular}}
        \vspace*{-0.5\baselineskip}
\end{table}

The online experimental results are shown in Table~\ref{online_performance_pv_click} and Table~\ref{online_performance_uv_ucvr}. Referring to the table, we can observe that the PV and Click metrics of the above scenes get a dramatic improvement compared with the base group, which means (1) the incremental categories recalled by the new model MMAN are indeed the category the users required; (2) by increasing the recall rate of related categories, users tend to view and click more products. With the increase in product selection, the conversion rate of users has increased a lot, leading to more GMV and UCVR improvement (+0.351\%).


\section{Conclusion and Future Work}
In this paper, we propose a multi-granularity matching attention network to comprehensively extract features from the char-level and semantic-level of a query-category interaction matrix. In this way, the model can overcome the problem of a lack of training samples for long-tail queries and eliminate the difference in expression between queries and categories. The offline and online A/B experiments achieve significant improvements over the state of the art. Furthermore, MMAN has already been deployed in production at the JD application and brings 
great commercial value, which confirms that MMAN is a practical and robust solution for large-scale query intent classification services.

In future work, we plan to explore utilizing external knowledge such as the taxonomic hierarchy of categories and product information to completely model the category representations for further enhancing the model's performance.

\bibliographystyle{ACM-Reference-Format}
\bibliography{main}


\begin{thebibliography}{21}


\ifx \showCODEN    \undefined \def \showCODEN     #1{\unskip}     \fi
\ifx \showDOI      \undefined \def \showDOI       #1{#1}\fi
\ifx \showISBNx    \undefined \def \showISBNx     #1{\unskip}     \fi
\ifx \showISBNxiii \undefined \def \showISBNxiii  #1{\unskip}     \fi
\ifx \showISSN     \undefined \def \showISSN      #1{\unskip}     \fi
\ifx \showLCCN     \undefined \def \showLCCN      #1{\unskip}     \fi
\ifx \shownote     \undefined \def \shownote      #1{#1}          \fi
\ifx \showarticletitle \undefined \def \showarticletitle #1{#1}   \fi
\ifx \showURL      \undefined \def \showURL       {\relax}        \fi
\providecommand\bibfield[2]{#2}
\providecommand\bibinfo[2]{#2}
\providecommand\natexlab[1]{#1}
\providecommand\showeprint[2][]{arXiv:#2}

\bibitem[Cai et~al\mbox{.}(2021)]%
        {cai2021slim}
\bibfield{author}{\bibinfo{person}{Fengyu Cai}, \bibinfo{person}{Wanhao Zhou},
  \bibinfo{person}{Fei Mi}, {and} \bibinfo{person}{Boi Faltings}.}
  \bibinfo{year}{2021}\natexlab{}.
\newblock \showarticletitle{SLIM: Explicit Slot-Intent Mapping with BERT for
  Joint Multi-Intent Detection and Slot Filling}.
\newblock \bibinfo{journal}{\emph{arXiv preprint arXiv:2108.11711}}
  (\bibinfo{year}{2021}).
\newblock


\bibitem[Chen et~al\mbox{.}(2019)]%
        {chen2019bert}
\bibfield{author}{\bibinfo{person}{Qian Chen}, \bibinfo{person}{Zhu Zhuo},
  {and} \bibinfo{person}{Wen Wang}.} \bibinfo{year}{2019}\natexlab{}.
\newblock \showarticletitle{Bert for joint intent classification and slot
  filling}.
\newblock \bibinfo{journal}{\emph{arXiv preprint arXiv:1902.10909}}
  (\bibinfo{year}{2019}).
\newblock


\bibitem[Hashemi et~al\mbox{.}(2016)]%
        {hashemi2016query}
\bibfield{author}{\bibinfo{person}{Homa~B Hashemi}, \bibinfo{person}{Amir
  Asiaee}, {and} \bibinfo{person}{Reiner Kraft}.}
  \bibinfo{year}{2016}\natexlab{}.
\newblock \showarticletitle{Query intent detection using convolutional neural
  networks}. In \bibinfo{booktitle}{\emph{International Conference on Web
  Search and Data Mining, Workshop on Query Understanding}}.
\newblock


\bibitem[Kenton and Toutanova(2019)]%
        {kenton2019bert}
\bibfield{author}{\bibinfo{person}{Jacob Devlin Ming-Wei~Chang Kenton} {and}
  \bibinfo{person}{Lee~Kristina Toutanova}.} \bibinfo{year}{2019}\natexlab{}.
\newblock \showarticletitle{BERT: Pre-training of Deep Bidirectional
  Transformers for Language Understanding}. In
  \bibinfo{booktitle}{\emph{Proceedings of NAACL-HLT}}.
  \bibinfo{pages}{4171--4186}.
\newblock


\bibitem[Kingma and Ba(2014)]%
        {kingma2014adam}
\bibfield{author}{\bibinfo{person}{Diederik~P Kingma} {and}
  \bibinfo{person}{Jimmy Ba}.} \bibinfo{year}{2014}\natexlab{}.
\newblock \showarticletitle{Adam: A method for stochastic optimization}.
\newblock \bibinfo{journal}{\emph{arXiv preprint arXiv:1412.6980}}
  (\bibinfo{year}{2014}).
\newblock


\bibitem[Lai et~al\mbox{.}(2015)]%
        {lai2015recurrent}
\bibfield{author}{\bibinfo{person}{Siwei Lai}, \bibinfo{person}{Liheng Xu},
  \bibinfo{person}{Kang Liu}, {and} \bibinfo{person}{Jun Zhao}.}
  \bibinfo{year}{2015}\natexlab{}.
\newblock \showarticletitle{Recurrent convolutional neural networks for text
  classification}. In \bibinfo{booktitle}{\emph{Twenty-ninth AAAI conference on
  artificial intelligence}}.
\newblock


\bibitem[Liu et~al\mbox{.}(2017)]%
        {liu2017deep}
\bibfield{author}{\bibinfo{person}{Jingzhou Liu}, \bibinfo{person}{Wei-Cheng
  Chang}, \bibinfo{person}{Yuexin Wu}, {and} \bibinfo{person}{Yiming Yang}.}
  \bibinfo{year}{2017}\natexlab{}.
\newblock \showarticletitle{Deep learning for extreme multi-label text
  classification}. In \bibinfo{booktitle}{\emph{Proceedings of the 40th
  international ACM SIGIR conference on research and development in information
  retrieval}}. \bibinfo{pages}{115--124}.
\newblock


\bibitem[Liu et~al\mbox{.}(2019)]%
        {liu2019roberta}
\bibfield{author}{\bibinfo{person}{Yinhan Liu}, \bibinfo{person}{Myle Ott},
  \bibinfo{person}{Naman Goyal}, \bibinfo{person}{Jingfei Du},
  \bibinfo{person}{Mandar Joshi}, \bibinfo{person}{Danqi Chen},
  \bibinfo{person}{Omer Levy}, \bibinfo{person}{Mike Lewis},
  \bibinfo{person}{Luke Zettlemoyer}, {and} \bibinfo{person}{Veselin
  Stoyanov}.} \bibinfo{year}{2019}\natexlab{}.
\newblock \showarticletitle{Roberta: A robustly optimized bert pretraining
  approach}.
\newblock \bibinfo{journal}{\emph{arXiv preprint arXiv:1907.11692}}
  (\bibinfo{year}{2019}).
\newblock


\bibitem[Mikolov et~al\mbox{.}(2013)]%
        {mikolov2013distributed}
\bibfield{author}{\bibinfo{person}{Tomas Mikolov}, \bibinfo{person}{Ilya
  Sutskever}, \bibinfo{person}{Kai Chen}, \bibinfo{person}{Greg~S Corrado},
  {and} \bibinfo{person}{Jeff Dean}.} \bibinfo{year}{2013}\natexlab{}.
\newblock \showarticletitle{Distributed representations of words and phrases
  and their compositionality}. In \bibinfo{booktitle}{\emph{Advances in neural
  information processing systems}}. \bibinfo{pages}{3111--3119}.
\newblock


\bibitem[Qiu et~al\mbox{.}(2022)]%
        {qiu2022pre}
\bibfield{author}{\bibinfo{person}{Yiming Qiu}, \bibinfo{person}{Chenyu Zhao},
  \bibinfo{person}{Han Zhang}, \bibinfo{person}{Jingwei Zhuo},
  \bibinfo{person}{Tianhao Li}, \bibinfo{person}{Xiaowei Zhang},
  \bibinfo{person}{Songlin Wang}, \bibinfo{person}{Sulong Xu},
  \bibinfo{person}{Bo Long}, {and} \bibinfo{person}{Wen-Yun Yang}.}
  \bibinfo{year}{2022}\natexlab{}.
\newblock \showarticletitle{Pre-training Tasks for User Intent Detection and
  Embedding Retrieval in E-commerce Search}. In
  \bibinfo{booktitle}{\emph{Proceedings of the 31st ACM International
  Conference on Information \& Knowledge Management}}.
  \bibinfo{pages}{4424--4428}.
\newblock


\bibitem[Sreelakshmi et~al\mbox{.}(2018)]%
        {sreelakshmi2018deep}
\bibfield{author}{\bibinfo{person}{K Sreelakshmi}, \bibinfo{person}{PC
  Rafeeque}, \bibinfo{person}{S Sreetha}, {and} \bibinfo{person}{ES Gayathri}.}
  \bibinfo{year}{2018}\natexlab{}.
\newblock \showarticletitle{Deep bi-directional LSTM network for query intent
  detection}.
\newblock \bibinfo{journal}{\emph{Procedia computer science}}
  \bibinfo{volume}{143} (\bibinfo{year}{2018}), \bibinfo{pages}{939--946}.
\newblock


\bibitem[Wang et~al\mbox{.}(2018)]%
        {wang2018joint}
\bibfield{author}{\bibinfo{person}{Guoyin Wang}, \bibinfo{person}{Chunyuan Li},
  \bibinfo{person}{Wenlin Wang}, \bibinfo{person}{Yizhe Zhang},
  \bibinfo{person}{Dinghan Shen}, \bibinfo{person}{Xinyuan Zhang},
  \bibinfo{person}{Ricardo Henao}, {and} \bibinfo{person}{Lawrence Carin}.}
  \bibinfo{year}{2018}\natexlab{}.
\newblock \showarticletitle{Joint Embedding of Words and Labels for Text
  Classification}. In \bibinfo{booktitle}{\emph{Proceedings of the 56th Annual
  Meeting of the Association for Computational Linguistics (Volume 1: Long
  Papers)}}. \bibinfo{pages}{2321--2331}.
\newblock


\bibitem[Wu et~al\mbox{.}(2019)]%
        {wu2019learning}
\bibfield{author}{\bibinfo{person}{Jiawei Wu}, \bibinfo{person}{Wenhan Xiong},
  {and} \bibinfo{person}{William~Yang Wang}.} \bibinfo{year}{2019}\natexlab{}.
\newblock \showarticletitle{Learning to learn and predict: A meta-learning
  approach for multi-label classification}.
\newblock \bibinfo{journal}{\emph{arXiv preprint arXiv:1909.04176}}
  (\bibinfo{year}{2019}).
\newblock


\bibitem[Xiao et~al\mbox{.}(2019)]%
        {xiao2019label}
\bibfield{author}{\bibinfo{person}{Lin Xiao}, \bibinfo{person}{Xin Huang},
  \bibinfo{person}{Boli Chen}, {and} \bibinfo{person}{Liping Jing}.}
  \bibinfo{year}{2019}\natexlab{}.
\newblock \showarticletitle{Label-specific document representation for
  multi-label text classification}. In \bibinfo{booktitle}{\emph{Proceedings of
  the 2019 conference on empirical methods in natural language processing and
  the 9th international joint conference on natural language processing
  (EMNLP-IJCNLP)}}. \bibinfo{pages}{466--475}.
\newblock


\bibitem[Yang et~al\mbox{.}(2016)]%
        {yang2016hierarchical}
\bibfield{author}{\bibinfo{person}{Zichao Yang}, \bibinfo{person}{Diyi Yang},
  \bibinfo{person}{Chris Dyer}, \bibinfo{person}{Xiaodong He},
  \bibinfo{person}{Alex Smola}, {and} \bibinfo{person}{Eduard Hovy}.}
  \bibinfo{year}{2016}\natexlab{}.
\newblock \showarticletitle{Hierarchical attention networks for document
  classification}. In \bibinfo{booktitle}{\emph{Proceedings of the 2016
  conference of the North American chapter of the association for computational
  linguistics: human language technologies}}. \bibinfo{pages}{1480--1489}.
\newblock


\bibitem[Yu et~al\mbox{.}(2018)]%
        {yu2018diverse}
\bibfield{author}{\bibinfo{person}{Mo Yu}, \bibinfo{person}{Xiaoxiao Guo},
  \bibinfo{person}{Jinfeng Yi}, \bibinfo{person}{Shiyu Chang},
  \bibinfo{person}{Saloni Potdar}, \bibinfo{person}{Yu Cheng},
  \bibinfo{person}{Gerald Tesauro}, \bibinfo{person}{Haoyu Wang}, {and}
  \bibinfo{person}{Bowen Zhou}.} \bibinfo{year}{2018}\natexlab{}.
\newblock \showarticletitle{Diverse few-shot text classification with multiple
  metrics}.
\newblock \bibinfo{journal}{\emph{arXiv preprint arXiv:1805.07513}}
  (\bibinfo{year}{2018}).
\newblock


\bibitem[Zhang et~al\mbox{.}(2019b)]%
        {zhang2019improving}
\bibfield{author}{\bibinfo{person}{Hongchun Zhang}, \bibinfo{person}{Tianyi
  Wang}, \bibinfo{person}{Xiaonan Meng}, \bibinfo{person}{Yi Hu}, {and}
  \bibinfo{person}{Hao Wang}.} \bibinfo{year}{2019}\natexlab{b}.
\newblock \showarticletitle{Improving Semantic Matching via Multi-Task Learning
  in E-Commerce.}. In \bibinfo{booktitle}{\emph{eCOM@ SIGIR}}.
\newblock


\bibitem[Zhang et~al\mbox{.}(2021)]%
        {zhang2021modeling}
\bibfield{author}{\bibinfo{person}{Junhao Zhang}, \bibinfo{person}{Weidi Xu},
  \bibinfo{person}{Jianhui Ji}, \bibinfo{person}{Xi Chen},
  \bibinfo{person}{Hongbo Deng}, {and} \bibinfo{person}{Keping Yang}.}
  \bibinfo{year}{2021}\natexlab{}.
\newblock \showarticletitle{Modeling Across-Context Attention For Long-Tail
  Query Classification in E-commerce}. In \bibinfo{booktitle}{\emph{Proceedings
  of the 14th ACM International Conference on Web Search and Data Mining}}.
  \bibinfo{pages}{58--66}.
\newblock


\bibitem[Zhang et~al\mbox{.}(2019a)]%
        {zhang2019ernie}
\bibfield{author}{\bibinfo{person}{Zhengyan Zhang}, \bibinfo{person}{Xu Han},
  \bibinfo{person}{Zhiyuan Liu}, \bibinfo{person}{Xin Jiang},
  \bibinfo{person}{Maosong Sun}, {and} \bibinfo{person}{Qun Liu}.}
  \bibinfo{year}{2019}\natexlab{a}.
\newblock \showarticletitle{ERNIE: Enhanced language representation with
  informative entities}.
\newblock \bibinfo{journal}{\emph{arXiv preprint arXiv:1905.07129}}
  (\bibinfo{year}{2019}).
\newblock


\bibitem[Zhao et~al\mbox{.}(2019)]%
        {zhao2019dynamic}
\bibfield{author}{\bibinfo{person}{Jiashu Zhao}, \bibinfo{person}{Hongshen
  Chen}, {and} \bibinfo{person}{Dawei Yin}.} \bibinfo{year}{2019}\natexlab{}.
\newblock \showarticletitle{A dynamic product-aware learning model for
  e-commerce query intent understanding}. In
  \bibinfo{booktitle}{\emph{Proceedings of the 28th ACM International
  Conference on Information and Knowledge Management}}.
  \bibinfo{pages}{1843--1852}.
\newblock


\bibitem[Zhou et~al\mbox{.}(2020)]%
        {zhou2020hierarchy}
\bibfield{author}{\bibinfo{person}{Jie Zhou}, \bibinfo{person}{Chunping Ma},
  \bibinfo{person}{Dingkun Long}, \bibinfo{person}{Guangwei Xu},
  \bibinfo{person}{Ning Ding}, \bibinfo{person}{Haoyu Zhang},
  \bibinfo{person}{Pengjun Xie}, {and} \bibinfo{person}{Gongshen Liu}.}
  \bibinfo{year}{2020}\natexlab{}.
\newblock \showarticletitle{Hierarchy-aware global model for hierarchical text
  classification}. In \bibinfo{booktitle}{\emph{Proceedings of the 58th Annual
  Meeting of the Association for Computational Linguistics}}.
  \bibinfo{pages}{1106--1117}.
\newblock


\end{thebibliography}


\end{document}